\documentclass[aps,prl,twocolumn,preprintnumbers]{revtex4-2}
\usepackage{amsmath, mathrsfs, amssymb,amsfonts,amsthm,graphicx, epsf, dcolumn, yfonts}
\usepackage[hyperfootnotes=true]{hyperref}
\usepackage{color}
\usepackage[usenames,dvipsnames]{xcolor}
\hypersetup{
  colorlinks   = true, %Colours links instead of ugly boxes
  urlcolor     = blue, %Colour for external hyperlinks
  linkcolor    = blue, %Colour of internal links
  citecolor   = red %Colour of citations
}

\usepackage{slashed}
\usepackage{tcolorbox}
\usepackage{setspace}
\usepackage{cancel}
\usepackage{wasysym}
\usepackage{amsmath}
\usepackage[lofdepth,lotdepth]{subfig}
\usepackage{float}
\usepackage{ragged2e}
\DeclareCaptionJustification{justified}{\justifying}
\pdfoutput=1
\begin{document}

\title{Spectral Density of the Causal Propagator}

 \author{Joshua Y. L. Jones$^{1,2}$}\email{jjones@stp.dias.ie} 
\author{Yasaman K. Yazdi$^1$}%
  \email{ykyazdi@stp.dias.ie}
\affiliation{$^1$ School of Theoretical Physics, Dublin Institute for Advanced Studies, 10 Burlington Road, Dublin 4, Ireland\\ $^2$ School of Mathematics, Trinity College Dublin,  20 Westland Row, 
Dublin 2, Ireland}

\date{\today}

\begin{abstract}
The causal propagator (or Pauli-Jordan function), which multiplied by $i$ is the spacetime commutator of the field $[\phi(x),\phi(x')]$, plays an essential role in scalar quantum field theory. We discuss the role of the causal propagator and its spectrum in recent developments in defining quantum field theory in a more explicitly covariant manner, as well as in causal set theory. We then present a conjecture for its asymptotic spectral density in a free theory, and give examples in Minkowski and de Sitter spacetimes that lend evidence to the conjectured scaling. Our work has implications for Lorentzian spectral geometry in much the same way as Weyl's asymptotic law has for Riemannian spectral geometry.
\end{abstract}

\maketitle

Spectral formulations often have the advantage of expressing physical quantities in an explicitly basis-independent manner. This is particularly useful and relevant in the context of classical and quantum gravity, where frame-independence (and therefore basis-independence) is necessary. 

In classical free scalar field theory, the causal propagator (or Pauli-Jordan function) $\Delta$ can be used to solve the Cauchy problem in globally hyperbolic spacetimes \cite{Fulling_1989}. It is defined as
\begin{equation}\label{eq: delta gr}
    \Delta(x,x'):=G_R(x,x')-G_A(x,x'),
\end{equation}
where $x$ and $x'$ are spacetime points (not necessarily at equal times) and $G_{R,A}$ are the retarded and advanced Green functions satisfying
\begin{equation}
    (\Box_x -m^2)G_{R,A}(x,x')=\frac{\delta^d(x-x')}{\sqrt{-g(x')}},
\end{equation}
where $d$ is the spacetime dimension. $G_R(x,x')$ is non-zero only if $x'$ causally precedes $x$, and  $G_A(x,x')=G_R(x',x)$. We also have \cite{Sorkin:2017fcp}
\begin{equation}
    \text{image}(i\Delta) = \text{ker}(\Box-m^2),
\end{equation}
which is not surprising given that $\Delta$ can be used to solve the Cauchy problem and hence must have full information of the solutions of the field equation. As an operator, $\Delta$ is also closely related to null geodesics, along which it propagates singularities \cite{KM2015}. In the quantum theory, $i\Delta$ is the commutator  of the field operator:
\begin{equation}\label{eq: commutator}
    i\Delta(x,x')=[\phi(x),\phi(x')].
\end{equation}
The \emph{Sorkin-Johnston} (SJ) prescription \cite{Johnston:2009fr, Sorkin:2017fcp} is an algebraic quantum field theory approach that takes cue from the large amount of information encoded in $i\Delta$, and defines a Wightman function $W(x,x')=\langle0|\phi(x)\phi(x')|0\rangle$, using its positive spectral part:
\begin{equation}\label{eq: WSJ}
    W_{SJ}(x,x'):=\text{pos}(i\Delta).
\end{equation}
This is always possible as $i\Delta$ is self-adjoint (or can be made to be) and anti-symmetric, ensuring that its nonzero eigenvalues come in $\pm$ real pairs. Explicitly, the eigenfunctions $\{f_\lambda\}$ and eigenvalues $\{\lambda\}$ of $i\Delta$ on a Lorentzian manifold $\mathcal M$ (which we take to be compact and globally hyperbolic) solve
\begin{equation}
    \int_{\mathcal M}i\Delta(x,x')f_\lambda(x')dV'=\lambda f_\lambda(x).
\end{equation}
In terms of these eigenfunctions and eigenvalues, \eqref{eq: WSJ} can be written more explicitly as
\begin{equation}
      W_{SJ}(x,x')=\sum_\lambda\lambda f_\lambda(x) f^*_\lambda(x');\qquad \lambda>0,
\end{equation}
with the $f_\lambda$'s being $L^2$-normalized. A nice feature of \eqref{eq: WSJ} is that it is defined spectrally and in terms of an operator that has support in \emph{spacetime},  making it explicitly frame-independent. This is in contrast to usual methods for defining a $W$ which rely on the existence of a symmetry or else some other preferred frame to select a subspace or state; in general curved spacetimes, there is no natural preferred frame.

The SJ prescription can also be used to define a free scalar field (via $W_{SJ}$) on a causal set (a fundamentally discrete structure proposed to underlie spacetime in the causal set theory approach to quantum gravity). The quantum field on a causal set has the added benefit that it is automatically ultraviolet regulated due to the discreteness. Moreover, it is easier to find the eigenvalues and eigenvectors of $i\Delta$ in this case since it is a finite-dimensional matrix. For more details on $W_{SJ}$, we refer the reader to \cite{Johnston:2010su, Afshordi:2012ez, Afshordi:2012jf, Jones:2024yzf}.

The spectrum of $i\Delta$ also plays an important role in a spectral spacetime formulation of entanglement entropy for free fields \cite{Sorkin:2012sn, Chen:2020ild, Saravani:2013nwa, Jones:2024yzf, Mathur:2021ial, Jones:2026vkw}. In this formulation, the entropy of a field in a Gaussian state $W$ and in a manifold $\mathcal M$ is
\begin{equation}
    S=\sum_{\tilde{\lambda}} \tilde{\lambda}\ln|\tilde{\lambda}|,
\end{equation}
where $\{\tilde{\lambda}\}$ are solutions to the generalized eigenvalue equation
\begin{equation} \label{eq: gen eig eq}
  \int_{\mathcal M}W(x,x')\tilde{f}(x')dV'=\tilde{\lambda}  \int_{\mathcal M}i\Delta(x,x')\tilde{f}(x')dV',
\end{equation}
provided we mod out the kernel of $\Delta$, and implement an ultraviolet cutoff before solving \eqref{eq: gen eig eq}. Since $W$ and $i\Delta$ are spacetime functions, they lend themselves to more explicitly covariant spacetime regularization schemes such as via the inherent discreteness of a causal set background (as done in \cite{Sorkin:2016pbz, Surya:2020gjj, Duffy:2021dtc, Keseman:2021dkf, Jones:2026vkw}) or via a truncation in the spectrum of $i\Delta$ (as done in \cite{Saravani:2013nwa}). Knowledge of the asymptotic properties of the spectrum of $i\Delta$ also facilitates a technical procedure in the causal set calculations whereby some spurious contributions (with no continuum analogue) near the discreteness scale need to be removed or \emph{truncated} to obtain meaningful results.

Thus, we can enumerate several broad reasons to be interested in the spectrum of the causal propagator $\Delta$ or spacetime commutator $i\Delta$: (1) It provides insight into the field's degrees of freedom in a basis-independent manner, (2) It is central to the SJ prescription for defining a Wightman function, (3) It can be used to covariantly implement a cutoff (e.g. for entanglement entropy), and (4) It is of general interest for Lorentzian spectral geometry.

\begin{figure}[t]
\centering
\includegraphics[width=.8\linewidth]{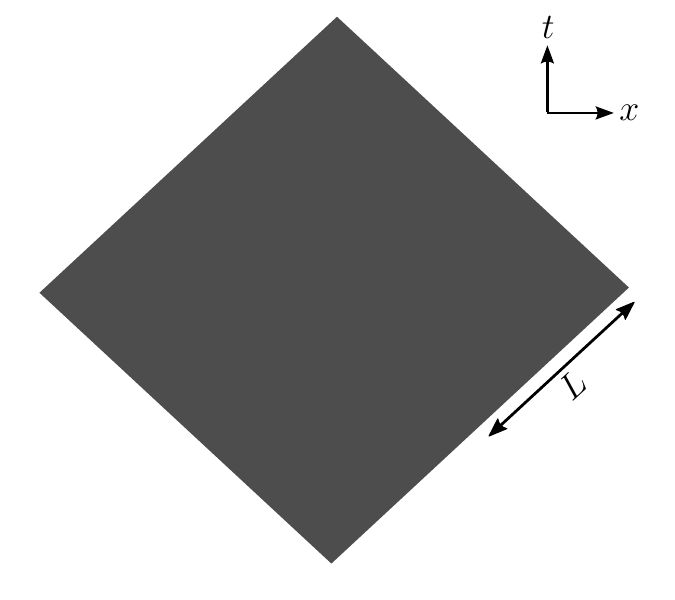}
\caption{\label{fig: diamond} A causal diamond in $1+1$d Minkowski spacetime.}
\end{figure}

\begin{figure}[b]
\centering
\includegraphics[width=.8\linewidth]{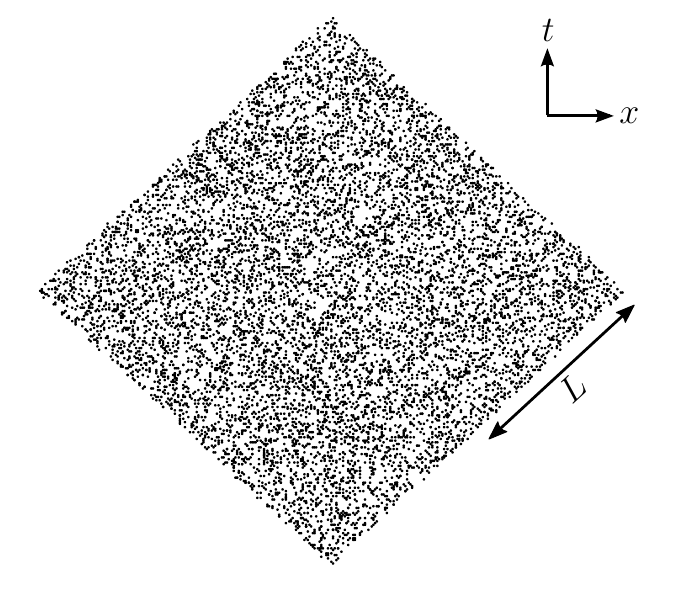}
\caption{\label{fig:cs diamond} A $10,000$-element causal set approximating a causal diamond in $1+1$d Minkowski spacetime, shown as an embedding in Minkowski spacetime. The elements are uniformly and randomly distributed, and the probability to find $N$ elements in a region with spacetime volume $V$ is given by the Poisson distribution,  $P_N(V) = \frac{(\rho V)^N}{N!}e^{-\rho V}$ where $\rho=\frac {\langle N\rangle} V$.}
\end{figure}

\begin{figure}[h]
\includegraphics[width=1.\linewidth]{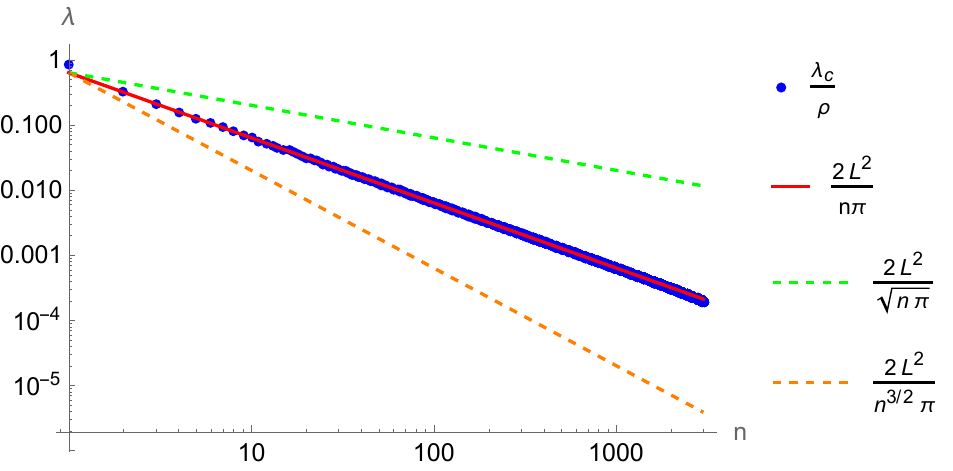}
\caption{\label{fig:2d scaling} Scaling of the largest 3000 positive eigenvalues of $i\Delta_c^{(2d)}$. For comparison, in addition to the expected inverse power-law scaling shown by the solid line, two additional power-law scalings are shown by the dashed lines. In this example, $\rho=\frac{\langle N\rangle}{V}=\frac{20000}{4L^2}$ and $L=1$. A rescaling factor of $1/\rho$ is needed to compare the dimensionless $\lambda_c$ with $\lambda$ which has dimensions of length squared.}
\end{figure}

Let us now review the well-studied \cite{Afshordi:2012ez} example of the spectrum of $i\Delta$ for a massless theory in a causal diamond in $1+1$-dimensional Minkowski spacetime. It is convenient to work in lightcone coordinates, where  $u = (t+\mathbf{x})/ \sqrt{2}$ and $v = (t-\mathbf{x})/ \sqrt{2}$. We choose $-L\leq u,v\leq L$; see Figure \ref{fig: diamond} for the setup. The causal propagator is \cite{Johnston:2010su, Egorov1994}
\begin{equation}\label{eq: Delta for massless 2d diamond}
    \Delta^{(2d)}(u,v;u',v')=\left(1-\Theta(u-u')-\Theta(v-v')\right)/2,
\end{equation}
and $i \Delta^{(2d)}$ has two families of eigenfunctions: 
\begin{align}
f_k(u,v):=& e^{-iku} - e^{-ikv} ; ~~k = \frac{n \pi }{L},~ n = \pm1,\pm2,..., \label{eq: f eig}\\
g_k(u,v):=& e^{-iku} + e^{-ikv} - 2\,\textrm{cos}(kL); ~~ k \in \mathcal{K}, \label{eq: g eig}
\end{align}
where $\mathcal{K} = \{ k \in \mathbb{R} ~|~ \textrm{tan}(kL) = 2kL  \}.$ The eigenvalues for both families of eigenfunctions are $\lambda_k=\frac{L}{k}$, with the $k$'s as specified in \eqref{eq: f eig} and \eqref{eq: g eig}, and asymptotically
\begin{equation}\label{eq: asymptotic spec 2d diamond}
    \lim_{n\rightarrow\infty}\lambda^{(2d)}_n\sim\frac{L^2}{n\pi}.
\end{equation}
We can independently verify this asymptotic scaling using a causal set diamond, depicted in Figure \ref{fig:cs diamond}. 
The causal propagator \eqref{eq: Delta for massless 2d diamond} in terms of causal set quantities is \cite{Johnston:2010su, Afshordi:2012ez}
\begin{equation} \label{eq: Delta cs for massless 2d diamond}
    \Delta^{(2d)}_c=\left(C^T-C\right)/2,
\end{equation}
 where $C$ is the causal matrix, which in a basis labelled by the elements, has entries
\begin{equation}
    C_{xx'}:=\begin{cases}
1, & \text{if $x\prec x'$}\\
0, & \text{otherwise},
\end{cases}
\end{equation}
 where $\prec$ denotes causal precedence. We have used the subscript $c$ in \eqref{eq: Delta cs for massless 2d diamond} to distinguish the causal set function from its continuum counterpart \eqref{eq: Delta for massless 2d diamond}. In Figure~\ref{fig:2d scaling}, the largest $3000$ positive eigenvalues of $i\Delta^{(2d)}_c$ for a $20,000$-element $1+1$d causal set diamond are shown  on a log-log scale. This range of the eigenvalues suffices to capture the power-law character of the spectrum. We do not include the smaller eigenvalues, because we would approach the discreteness scale and enter a regime with  contributions which do not have a continuum analogue (see e.g. \cite{Sorkin:2016pbz, Keseman:2021dkf, Jones:2026vkw}). This occurs at roughly $\lambda^{(2d)}_{\rho}\sim\frac{L}{2 \pi\sqrt{\rho}}$, which in our example is $\sim 0.002$. We can see that the scaling agrees with \eqref{eq: asymptotic spec 2d diamond}, with a factor of $2$ to account for the approximate two-fold degeneracy (from the $\{f_k\}$ and $\{g_k\}$ families).
From \eqref{eq: asymptotic spec 2d diamond}, and accounting for the degeneracy, we can calculate the spectral density:
\begin{equation}\label{eq: spec density 2d diamond}
  \left  |\frac{d n}{d\lambda}\right|\sim\frac{n^2 \pi}{2L^2}\sim\frac{2L^2}{\pi\lambda^2}.
\end{equation}

We will state a conjecture for this spectral density in the ultraviolet (small eigenvalue) limit in a general spacetime. Before doing so, we  introduce some objects.

We can form a $2(d-1)$-dimensional phase space, the symplectic space of null geodesics with momentum, $\mathcal{N}_{k}$, by starting with the cotangent bundle of the manifold $\mathcal M$, imposing the null condition on the momenta, and then quotienting by the geodesic flow.

The symplectic form there is given by $\omega = d\theta$, where $\theta = k_\mu dx^{\mu}|_{\mathcal{N}_k}$ is the one-form on $\mathcal{N}_{k}$ descended to by the tautological one-form on the cotangent bundle. We write the Liouville measure, $\Omega=\frac{\omega^{d-1}}{(d-1)!}=d^{d-1}x\,d^{d-1}k$, and then pass to a semiclassical phase space (with $\hbar=1$) with the usual Planckian minimum volume, and measure $\tilde{\Omega} = \frac{\Omega}{(2\pi)^{d-1}} =  \frac{d^{d-1}x\,d^{d-1}k}{(2\pi)^{d-1}}$.

The conjecture is then as follows:
\begin{tcolorbox}\textbf{Conjecture:} \emph{
Each eigenfunction of $i\Delta$ is asymptotically associated with a semiclassical Liouville volume, which corresponds to a null geodesic $\gamma$ with momentum $k$. We thus identify the spectral measure of $i\Delta$ on the manifold $\mathcal{M}$ with the semiclassical Liouville measure of $\mathcal{N}_{k}$, our phase space: $\tilde{\rho}(\lambda)d\lambda\sim\frac{d^{d-1}x\,d^{d-1}k}{(2\pi)^{d-1}}$, where $\tilde{\rho}(\lambda):= \left|\frac{dn}{d\lambda}\right|$. Each eigenvalue $\lambda$ has the form $\frac{1}{2}\int_{\gamma}d\alpha_k$, where $\alpha_k$ is the affine parameter that gives the momentum (on phase space) as the tangent vector (on spacetime).}\end{tcolorbox}

We can clarify the conjecture by writing it more explicitly. We consider $\lambda>0$ and write the counting function
\begin{equation}\label{eq: countfn}
    n(\lambda) = \int_{\mathcal{N}_{k}} \Theta(\lambda'(\gamma_{k})-\lambda) \,\tilde{\Omega},
\end{equation}
where $\lambda'$ is the eigenvalue function, $\lambda':\mathcal{N}_{k}\to\mathbb{R}$, $\gamma_{k}\mapsto \frac{1}{2}\int_{\gamma}d\alpha_k$. We have the condition $d\alpha(k^\sharp)=1$, where $k^\sharp =g^{-1}(k,\cdot)$ is the tangent vector to $\gamma$, from Hamilton's equations. Equation (\ref{eq: countfn}) counts the volumes of phase space which map to an eigenvalue $\lambda'$ greater than the $\lambda$ specified.

We can then get the spectral density $\tilde{\rho}(\lambda)$ as
\begin{equation}
     \left  |\frac{dn}{d\lambda}\right| = \int_{\mathcal{N}_{k}}\delta(\lambda-\lambda')\,\tilde{\Omega} = \int_{\mathcal{N}_{k}}\delta\left(\lambda-\frac{1}{2}\int_{\gamma}d\alpha_k\right)\,\tilde{\Omega},
\end{equation}
where we have now inserted the eigenvalue map explicitly. 
Choosing a Cauchy surface $\Sigma$ that each geodesic intersects once, and taking coordinates, we have
\begin{equation}
   \left  |\frac{dn}{d\lambda}\right| = \int_{\Sigma}\delta\left(\lambda-\frac{1}{2}\int_{\gamma}\frac{dx^{\mu}}{g^{\mu \nu} k_\nu}\right)\frac{d^{d-1}x\,d^{d-1}k}{(2\pi)^{d-1}},
\end{equation}
where $g^{\mu\nu}k_{\nu}=\frac{dx^\mu}{d\alpha_k}$ is the tangent vector in coordinates, and there is no contraction over $\mu$. As written, the eigenvalues and measure are hypersurface-independent, and diffeomorphism invariant.

Let us check $\tilde{\rho}(\lambda)$ for the $1+1$d diamond we have just reviewed. We find:

\begin{equation}\label{eq: spec density 2d diamond conjecture}
    \begin{split}
      \left|\frac{d n}{d\lambda}\right|&=\int dx\int \frac{dk}{2\pi} \delta\left(\lambda-\frac{\Delta t}{2|k|}\right)\\
        &=\frac{2R}{2\pi}\int dk\,\delta\left(\lambda-\frac{R}{2|k|}\right)\\
        &=2\times\frac{2R}{2\pi} \frac{2 k^2}{R}\Big|_{k=\frac{R}{2\lambda}}=\frac{R^2}{\pi\lambda^2}=\frac{2L^2}{\pi\lambda^2},
    \end{split}
\end{equation}
where we integrated over the corner to corner middle hypersurface of the diamond, which has length $2R=2\sqrt{2}L$. In this calculation, and those following, we use the fact that we only need one coordinate component to calculate $\int_{\gamma}d\alpha_k$. We choose the $t$ coordinate, and then the fact that $\frac{dt}{d\alpha_k}=k^{\sharp}_{t}=|k|$ is constant allows us to write $\int_{\gamma}d\alpha_k=\frac{\Delta t}{|k|}$, where $\Delta t$ is the elapsed coordinate time on the geodesic inside the diamond. In the last line we have used that $\delta(f(x))=\frac{\delta(x-x_0)}{|f'(x_0)|}$,  where $x_0$ is the root of $f$. We see that \eqref{eq: spec density 2d diamond} and \eqref{eq: spec density 2d diamond conjecture} agree.

One may also interpret $\lambda$ as the area swept out by a wavelength corresponding to the momentum on phase space, traversing the length of the geodesic.

Next we compute $\tilde{\rho}(\lambda)$ for a $2+1$d Minkowski diamond:
\begin{equation}\label{eq: spec density 3d diamond conjecture}
    \begin{split}
        \left  |\frac{d n}{d\lambda}\right|&=\int d^2x\int \frac{d^2k}{(2\pi)^2} \delta\left(\lambda-\frac{\Delta t(\vec{x},\hat{k})}{2|k|}\right)\\
        &=\int d^2x\int \frac{d^2k}{(2\pi)^2} \delta\left(\lambda-\frac{1}{2|k|} \frac{R(R^{2}-|\vec{x}^{2}|)}{(R^{2}-(\vec{x}\cdot\hat{k})^{2})}\right)\\
        &=\frac{R^4}{15\lambda^3}=\frac{4 L^4}{15\lambda^3},
    \end{split}
\end{equation}
where the surface integrated over was the middle disk $\Sigma$, which has radius $R=\sqrt{2}L$, and $\Delta t(\vec{x},\hat{k})$ is the coordinate time elapsed inside the diamond for a geodesic that intersects $\Sigma$ at $\vec{x}$, with tangent direction $\hat{k}$.

We see the positive eigenvalues asymptotically scale as
\begin{equation}\label{eq: 3d scaling}
     \lim_{n\rightarrow\infty}\lambda^{(3d)}_n\sim\sqrt{\frac{2}{15}}\frac{L^2}{\sqrt{n}}.
\end{equation}
The retarded Green function (from which $\Delta$ can be obtained via \eqref{eq: delta gr}) for a massless scalar field in $2+1$d Minkowski spacetime is \cite{Johnston:2010su, Egorov1994}
\begin{equation}\label{eq: gr in 3d cont}
    G_R^{(3d)}(x,x')=\Theta(t-t')\Theta(\tau^2(x,x'))\frac{1}{2\pi\, \tau(x,x')},
\end{equation}
where $\tau(x,x'):=\sqrt{(t-t')^2-(|\vec{x}|-|\vec{x}\,'|)^2}$ is the  proper time between $x$ and $x'$. The eigenvalues and eigenfunctions of $\Delta$ in higher dimensional diamonds than $1+1$ are not known analytically. Therefore we will use a causal set to numerically study the spectrum. 

In a $2+1$d Minkowski causal set, $\tau(x,x')$ is (up to an approximately known proportionality constant) approximated by the length of the longest chain (a sequence of totally ordered elements) between elements $x$ and $x'$ \cite{PhysRevLett.66.260}. 

\begin{figure}[h]
\includegraphics[width=1\linewidth]{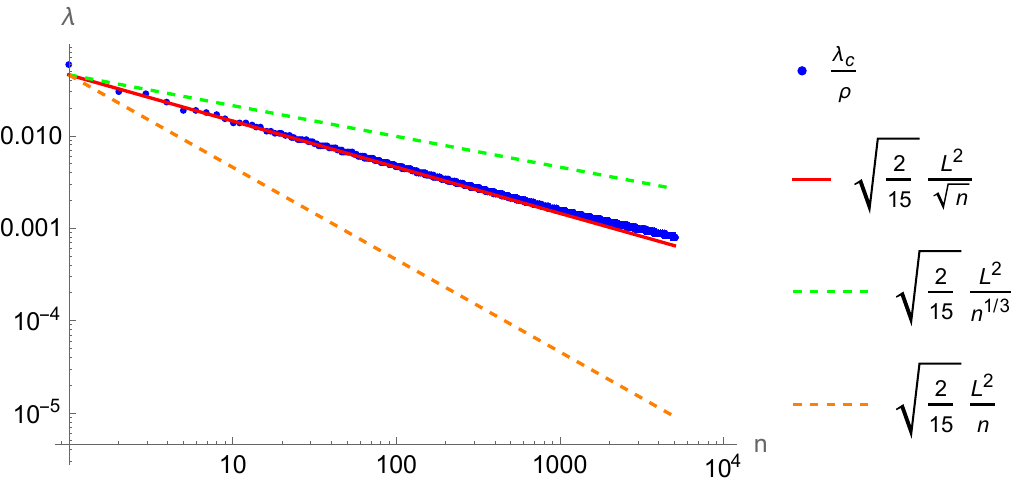}
\caption{\label{fig:3d scaling} Scaling of the largest 5000 positive eigenvalues of $i\Delta_c^{(3d)}$. In this example, $\rho=\frac{\langle N\rangle}{V}=\frac{80000}{\frac{4}{3} \sqrt{2} \pi  L^3}$ and $L=\frac{1}{\sqrt{8}}$. A rescaling factor of $1/\rho$ is needed to compare $\lambda_c$ which has length dimensions $-1$ with $\lambda$ which has dimensions of length squared.}
\end{figure}

In Figure \ref{fig:3d scaling}, the largest $5000$ positive eigenvalues of $i\Delta_c^{(3d)}$ in a $80,000$-element $2+1$d causal set diamond are shown on a log-log scale. Again we exclude the eigenvalues below the discreteness scale, roughly at $\lambda^{(3d)}_{\rho}\sim\frac{L}{2 \pi(\rho)^{1/3}}$, which in our example is $\sim 0.0008$. We see that the asymptotic scaling agrees with \eqref{eq: 3d scaling}.

Similarly, for a $3+1$d Minkowski causal diamond:
\begin{equation}\label{eq: spec density 4d diamond conjecture}
    \begin{split}
        \left  |\frac{d n}{d\lambda}\right|&=\int d^3x\int \frac{d^3k}{(2\pi)^3} \delta\left(\lambda-\frac{\Delta t(\vec{x},\hat{k})}{2|k|}\right)\\
        &=\int d^3x\int \frac{d^3k}{(2\pi)^3} \delta\left(\lambda-\frac{1}{2|k|} \frac{R(R^{2}-|\vec{x}^{2}|)}{(R^{2}-(\vec{x}\cdot\hat{k})^{2})}\right)\\
        &=\frac{R^6}{48\pi\lambda^4}=\frac{L^6}{6\pi\lambda^4},
    \end{split}
\end{equation}
where, as in the previous cases, the integration is over the (3d) middle sphere, which has radius $R=\sqrt{2}L$, and the intermediate objects are as previously defined. From \eqref{eq: spec density 4d diamond conjecture} we have for the positive eigenvalues that
\begin{equation}\label{eq: 4d scaling}
     \lim_{n\rightarrow\infty}\lambda^{(4d)}_n\sim \frac{L^2}{2} \left(\frac{4}{9\pi n}\right)^{1/3}.
\end{equation}

The retarded Green function for a massless scalar field in $3+1$d Minkowski spacetime is \cite{Johnston:2010su, Egorov1994}
\begin{equation}\label{eq: gr in 4d}
    G_R^{(4d)}(x,x')=\Theta(t-t')\Theta(\tau^2(x,x'))\frac{1}{2\pi}\delta(\tau^2(x,x')).
\end{equation}
The causal set analogue of \eqref{eq: gr in 4d} is  \cite{Johnston:2010su}
\begin{equation} \label{eq: gr cs for massless 4d}
    (G_R^{(4d)})_c=\frac{\sqrt{\rho}}{2\pi\sqrt{6}}\tilde{L}^T,
\end{equation}
 where $\tilde{L}$ is the link matrix
\begin{equation}
  \tilde{  L}_{xx'}:=\begin{cases}
1, & \text{if $x\prec x'\, \wedge\, \nexists\, y\,|\,x\prec y\prec x'$}\\
0, & \text{otherwise}.
\end{cases}
\end{equation}

\begin{figure}[h]
\includegraphics[width=1.\linewidth]{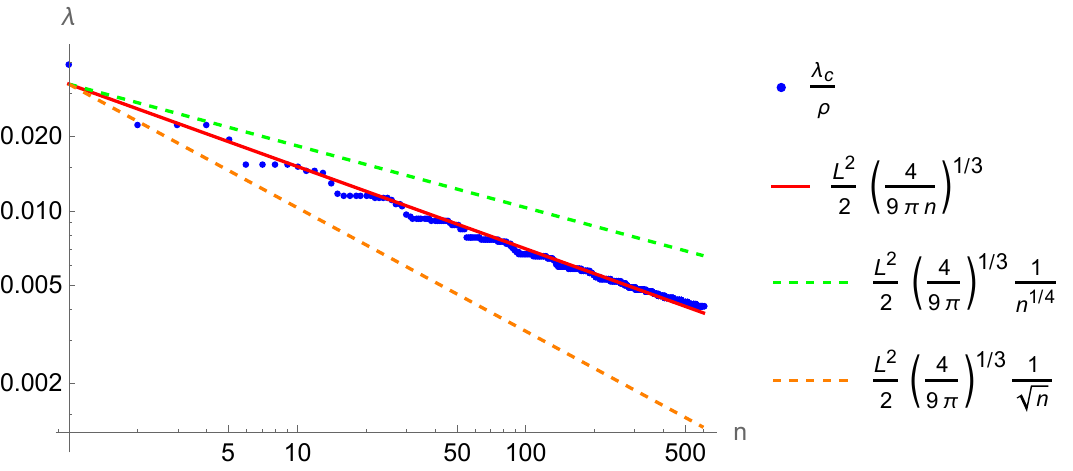}
\caption{\label{fig:4d scaling} Scaling of the largest 600 positive eigenvalues of $i\Delta_c^{(4d)}$. 
In this example, $\rho=\frac{\langle N\rangle}{V}=\frac{200000}{\frac{8}{3}\pi L^4}$ and $L=\frac{1}{\sqrt{8}}$. A rescaling factor of $1/\rho$ is needed to compare $\lambda_c$ which has length dimensions $-2$ with $\lambda$ which has dimensions of length squared.}
\end{figure}

It is computationally challenging to evaluate the spectrum of $i\Delta_c^{(4d)}$ for large enough causal sets that exhibit the asymptotic scaling without going over to the non-continuumlike regime. The latter occurs roughly at $\lambda^{(4d)}_{\rho}\sim\frac{L}{2 \pi(\rho)^{1/4}}$, which in our example is $\sim 0.002$. Compounded on this is the fact that \eqref{eq: gr cs for massless 4d} only becomes a good approximation to the delta function in \eqref{eq: gr in 4d} for very very large causal sets. Nevertheless, in Figure \ref{fig:4d scaling} we show the largest $600$ positive eigenvalues of $i\Delta_c^{(4d)}$ in a $200,000$-element causal set diamond on a log-log scale. The scaling roughly agrees with \eqref{eq: 4d scaling}. 

As a final test of our conjecture, we will consider an example in curved spacetime. We will calculate the spectral density of the causal propagator in a slab of de Sitter spacetime in $d$ dimensions. It has metric
\begin{equation}
ds^{2} = \left( \frac{1}{H\cos (\eta)} \right)^{2} (-d \eta^{2} +d\Omega^{2}_{d-1}),
\end{equation}
with time coordinate $\eta\in[-\tau,\tau]$ and de Sitter radius $H^{-1}$. 
This spacetime is conformal to the Einstein cylinder, which is an ultrastatic spacetime, for which methods are known to calculate the spectrum of the causal propagator \cite{Fewster}. Using this fact, the distinct eigenvalues in the de Sitter slab of a conformally coupled massless scalar field have been found to be \cite{Mathur}
\begin{gather}\label{eq: exactlambda}
\lambda_{n} = \pm \frac{1}{H^{2}\omega_{n}}\sqrt{ \int_{-\tau}^{\tau}d \eta' \frac{\sin^{2}(\omega_{n}\eta')}{\cos ^{2}(\eta')} \int_{-\tau}^{\tau}d \eta'' \frac{\cos^{2}(\omega_{n}\eta'')}{\cos ^{2}(\eta'')}}, \\
\omega_{n}=n+\frac{d-2}{2}, \quad n = 0,1,2,\ldots
\end{gather}
with degeneracy
\begin{equation}
\frac{(2n+d-2)(n+d-3)!}{n!(d-2)!}.
\end{equation}
In the asymptotic limit, accounting for the degeneracy,  \eqref{eq: exactlambda} gives a spectral density
\begin{equation}
\left|\frac{dn}{d\lambda}\right| \sim \frac{2}{(d-2)!}\left( \frac{\tan(\tau)}{H^{2}} \right)^{d-1} \frac{1}{\lambda^{d}}.
\end{equation}

We can now calculate this spectral density using the conjecture. Taking the $(d-1)$-sphere at $\eta=0$ as our hypersurface, with $\tilde{g}_{ij}$ as its induced metric, we calculate
\begin{equation}
\begin{aligned}
\left| \frac{dn}{d\lambda} \right| &= \int_{S^{d-1}} d^{d-1}x\int\frac{\,d^{d-1}k}{(2\pi)^{d-1}} \delta\left( \lambda - \frac{\tan(\tau)}{H^{2}\sqrt{ \tilde{g}^{ij}k_{i}k_{j} }} \right) \\
&= \frac{|S^{d-1}| \,|S^{d-2}|}{(2\pi)^{d-1}} \int_{0}^{\infty} dk\ k^{d-2} \delta\left(  \lambda - \frac{\tan(\tau)}{H^{2} k}\right) \\
%&= \frac{|S^{d-1}| \,|S^{d-2}|}{(2\pi)^{d-1}} \left( \frac{\tan(\tau)}{H^{2}} \right)^{d-1} \frac{1}{\lambda^{d}} \\
&= \frac{2}{(d-2)!} \left( \frac{\tan(\tau)}{H^{2}} \right)^{d-1} \frac{1}{\lambda^{d}},
\end{aligned}
\end{equation}
exactly as expected.

Our conjecture provides the Lorentzian counterpart of Weyl's law \cite{Weyl1911, 1969}, which describes the asymptotic density of the eigenvalues $\bar\lambda$ of the Laplacian in a Riemannian manifold. In a compact manifold $M$, this relation is:
\begin{equation}
 \lim_{\bar\lambda\rightarrow\infty}   \left|\frac{dn}{d\bar\lambda}\right|=\frac{d}{2(2\pi)^d}\bar\lambda^{d/2-1}B_d\text{Vol}(M),
\end{equation}
where $B_d$ is the volume of the unit ball in $\mathbb R^d$.
Weyl's asymptotic law has had many important applications in physics and mathematics \cite{Ivrii_2016}. Some areas of application include non-commutative geometry \cite{Fathizadeh:2013vp, ponge2021connesintegrationweylslaws}, Euclidean path integrals and Euclidean quantum gravity \cite{PhysRevD.18.1747, Reitz:2023ezz, Panine:2016pje}. Since spacetime is a Lorentzian manifold, a Lorentzian analogue of Weyl's law, such as the one we provide, will have useful applications in quantum gravity. \\

\begin{acknowledgments}
We would like to thank Rafael Sorkin for proposing this project and for many insightful discussions.
 This work was conducted with the support of Research Ireland under grant number 22/PATH-S/10704. We are grateful for the hospitality of Perimeter Institute where part of this work was carried out. Research at Perimeter Institute is supported in part by the Government of Canada through the Department of Innovation, Science and Economic Development and by the Province of Ontario through the Ministry of Colleges and Universities. This work was supported by a grant from the Simons Foundation (1034867, Dittrich).
\end{acknowledgments}

\bibliography{apssamp}

\end{document}